\documentclass[%
 aip,
 amsmath,amssymb,
 reprint,%
]{revtex4-1}

\usepackage{graphicx}
\usepackage{dcolumn}
\usepackage{bm}

\usepackage[utf8]{inputenc}
\usepackage[T1]{fontenc}
\usepackage{mathptmx}
\usepackage{etoolbox}

\usepackage{xcolor}
\definecolor{darkgreen}{RGB}{0, 100, 0} 
\usepackage[normalem]{ulem} 

\usepackage{hyperref}

\usepackage{multirow}

\newcommand{\vmes}{{\bf v}}

\newcommand{\kbt}{k_BT}
\newcommand{\fij}{F_{ij}}

\newcommand{\rbij}{{\bf r}_{ij}}

\newcommand{\Aij}{A_{ij}}
\newcommand{\Bij}{B_{ij}}
\newcommand{\rhoi}{\rho_{i}}
\newcommand{\rhoj}{\rho_{j}}

\newcommand{\eij}{{{\bf e}_{ij}}}

\newcommand{\wbbarij}{{\bf \overline{W}}_{ij}}
\newcommand{\wbij}{{\bf {W}}_{ij}}
\newcommand{\fa}{a}
\newcommand{\fb}{b}

\newcommand{\der}{\text{d}}
\newcommand{\dime}{\mathcal{D}}

\newcommand{\ca}{C^{\alpha}}
\newcommand{\mua}{\mu^{\alpha}}
\newcommand{\sa}{S^{\alpha}}

\newcommand{\ulen}{l}
\newcommand{\utime}{\tau}
\newcommand{\umass}{m}
\newcommand{\uener}{\epsilon}

\makeatletter
\def\@email#1#2{%
 \endgroup
 \patchcmd{\titleblock@produce}
  {\frontmatter@RRAPformat}
  {\frontmatter@RRAPformat{\produce@RRAP{*#1\href{mailto:#2}{#2}}}\frontmatter@RRAPformat}
  {}{}
}%
\makeatother
\begin{document}

\preprint{AIP/123-QED}

\title[Smoothed Dissipative Particle Dynamics for Mesoscale Advection-Diffusion-Reaction Problems]{Smoothed Dissipative Particle Dynamics for Mesoscale Advection-\\Diffusion-Reaction Problems}
\author{Marina Echeverr\'ia Ferrero}
\email{mecheverria@bcamath.org }
\affiliation{Basque Center for Applied Mathematics, Alameda de Mazarredo 14, 48009. Bilbao, Spain
}%
\affiliation{University of the Basque Country UPV/EHU, Barrio de Sarriena s/n 48940. Leioa, Spain}

\author{Nicolas Moreno}%
\email{nmoreno@bcamath.org}
\affiliation{Basque Center for Applied Mathematics, Alameda de Mazarredo 14, 48009. Bilbao, Spain
}%

\author{Marco Ellero}
\affiliation{Basque Center for Applied Mathematics, Alameda de Mazarredo 14, 48009. Bilbao, Spain
}%
\affiliation{IKERBASQUE Basque Foundation for Science, Mar\'ia D\'iaz de Haro 3, 48013, Bilbao, Spain.}
\affiliation{Complex Fluids Research Group, Department of Chemical Engineering, Swansea University, SA1 8EN Swansea, UK.}

\date{\today}

\begin{abstract}
Smoothed dissipative particle dynamics (\textcolor{black}{SDPD}) is a widely used particle-based method for modeling soft matter systems at mesoscopic and macroscopic scales, offering thermodynamic consistency and direct control over the fluid’s transport properties. Here, we present an \textcolor{black}{SDPD} model that incorporates the transport of reactants on scales smaller than the discretizing particles, including the evolution of compositional fields. The proposed methodology is well-suited for modeling complex systems governed by advection-diffusion-reaction (ADR) dynamics. Implemented in LAMMPS, the model is validated using a range of benchmark problems spanning diffusion-dominated, reaction-dominated, and coupled ADR regimes. Our simulation results demonstrate that the implemented \textcolor{black}{SDPD} model effectively captures complex behaviors, such as Turing pattern formation. The proposed model holds promise for applications across various fields, including biology, chemistry, materials science, and environmental engineering.
\end{abstract}

\maketitle





\section{Introduction}

Advection-diffusion-reaction (ADR) problems are ubiquitous in many fields of science and engineering, including biology, chemistry, and environmental sciences. The ADR problems are characterized by the transport of species by a fluid flow, the diffusion of the species in the fluid, and the reaction occurring among species or surrounding domains. ADR problems are typically described by a set of partial differential equations (PDEs) that account for the evolution of the species concentration, the fluid velocity, and the fluid pressure. In general, systems where advection, diffusion, and reaction processes take place concurrently can be challenging to solve due to the nonlinear coupling between fields. The complexity of these systems can be further enhanced when additional effects associated with the temporal and spatial scales of the systems, such as thermal fluctuations, are also relevant. In biology, for instance, many of the processes within cells require the consideration of thermal fluctuations that can drive the systems in and out of equilibrium.\cite{gonzalez2002inflationary,ruel1999jensen,slein2023} Those biological processes are often driven by the interactions affected by local concentration of the species. The complex interplay between species transport and reaction, along with phase separation within organelles has been recognized as a critical aspect in several cellular function.\cite{boeynaems2018,zhao2020}

Currently, a variety of numerical methods exist to model Advection-Diffusion-Reaction (ADR) problems, including both grid-based approaches such as finite difference, finite element, finite volume methods, and lattice Boltzmann methods (LBM),\cite{kruger2017lattice} as well as particle-based methods like Smoothed Particle Hydrodynamics (SPH)\cite{cleary2021,sigalotti2025smoothed} and Dissipative Particle Dynamics (DPD).\cite{liu2015review,feng2020review} These methods can be applied at different levels of description depending on the scale of interest. At the microscale, characterized by high spatial and temporal resolution, Molecular Dynamics (MD) simulations with potentials rooted in classical mechanics are widely used. At the macroscale, the continuum approach predominates, typically described by the Navier-Stokes equations. For mesoscale phenomena, thermally driven flows are modeled using the framework of ``fluctuating hydrodynamics'' (FHD), also known as the Landau-Lifshitz-Navier-Stokes (LLNS) equations.\cite{bell2007,landau1987course} This framework extends traditional Navier-Stokes equations by incorporating random stress fluctuations correlated in space and time, as determined by the fluctuation-dissipation theorem (FDT), thus offering a refined description of fluid behavior at intermediate scales. \textcolor{black}{Numerical methods to solve fluctuating hydrodynamics have been proposed using both grid-based and particle-based techniques. Grid-based methods\cite{donev2010,balboa2012,delgadobuscalioni2003,usabiaga2014} directly discretize the Landau–Lifshitz Navier–Stokes equations, while particle-based methods such as DPD\cite{bian2015multi} and SDPD\cite{Espanol2003} provide a thermodynamically consistent version of fluctuating hydrodynamics with direct control over transport coefficients.}

Lagrangian stochastic particle descriptions are particularly useful for mesoscale phenomena due to their ability to reproduce hydrodynamic behavior across time and length scales beyond what is feasible in fully resolved MD simulations. For ADR problems, combining the reaction and diffusion of chemical species within particle-based fluid dynamic frameworks provides a powerful tool for modeling complex dynamics.\cite{drawert2016stoch,tartakovsky2007} Mesoscale methodologies for ADR problems have been proposed in the context of DPD, accounting for the transport of species.\cite{Li2015, guan2019} However, some relevant limitations of DPD include the lack of direct linkage between model and physical parameters, necessitating empirical, case-specific calibration.\cite{xu2019} Additionally, it struggles to reproduce high Schmidt numbers\cite{yaghoubi2015,santo2021} and cannot directly incorporate equations of state.

In contrast, Smoothed Dissipative Particle Dynamics (\textcolor{black}{SDPD})\cite{Espanol2003, ellero2018everything} is a particle-based method widely used to model complex fluids that combines the thermal fluctuations of DPD with discretization of the Navier-Stokes equations in Lagrangian form, as seen in SPH. This allows \textcolor{black}{SDPD} to consistently discretize the fluctuating Navier-Stokes equations and directly specify transport properties such as viscosity. \textcolor{black}{The Lagrangian description of SDPD facilitates the modeling of systems with complex geometries,\cite{litvinov2010splitting,cai2023arbitrary}} making it suitable for simulating a wide range of problems in both synthetic and biological contexts (including blood flow\cite{Moreno2013, Fedosov2014}, viruses\cite{moreno-chaparro2025}, polymer solutions\cite{Litvinov2016}, colloidal suspensions\cite{bian2012multiscale}, gel formation\cite{Zohravi2023, Zohravi2024}, and multiphase systems\cite{espinosamoreno2025}) where reaction–diffusion dynamics span micro- to macroscales.\cite{Kulkarni2013, muller2014, ellero2018everything, Moreno2023} 

In the context of ADR problems, \textcolor{black}{SDPD} schemes have been proposed by Drawert et al.\cite{Drawert2019} leveraging a spatial stochastic simulation algorithm for molecular-level dynamics and a deterministic approach for broader scales to optimize computational efficiency. Originally, Ellero et al.\cite{Ellero2013} used the GENERIC formalism to derive a discretized advection-diffusion equation for systems of Hookean dumbbells in solvent. A similar approach has been formulated by Petsev and coauthors\cite{Petsev2016} to implement two-component mixtures in \textcolor{black}{SDPD}, using the discretization of the advection-diffusion equation with thermal noise in the concentration field. {In contrast to these specialized applications, our formulation targets generic reactive scalar species with adjustable diffusivities and reaction kinetics. This generalization extends ADR-SDPD to a broader range of transport–reaction problems, while retaining thermodynamic consistency and direct tuning of transport properties}

In this work, we present a novel \textcolor{black}{SDPD} model for simulating ADR problems at the mesoscale, where species diffusivity can be directly addressed without the need for estimations. We implement the ADR-\textcolor{black}{SDPD} scheme in the open-source Large-scale Atomic/Molecular Massively Parallel Simulator (LAMMPS)\cite{Thompson2022lammps} (29 Oct. 2020 stable release), ensuring the code scalability, flexibility to model diverse systems, and seamless integration with other LAMMPS functionalities for molecular and rigid-body objects.

The paper is organized as follows. In Section \ref{sec2:sdpd-formulation}, we present the \textcolor{black}{SDPD} model for ADR problems. In Section \ref{sec:results}, we present the results for a variety of benchmark problems to validate the implemented ADR-\textcolor{black}{SDPD} model. Initially, we focus on the validation of diffusion-dominated and reaction-dominated systems separately. Subsequently, we tackle more complex scenarios, such as Turing pattern formation, demonstrating the model's capability to simulate systems with intertwined advection, diffusion, and reaction phenomena, using benchmarks grounded in frameworks for natural and engineered pattern formation. Finally, in Section \ref{sec:discussion}, we present the conclusions and future work.
 
\section{Compositional Smoothed Dissipative Particle Dynamics}\label{sec2:sdpd-formulation}

Considering a Lagrangian description of the fluid, the fluid-mass balance can be expressed as \( {\der \rho}/{\der t} = -\rho \nabla \cdot \vmes \), whereas the momentum equation is given by
\begin{gather}
\rho\frac{\der \vmes}{\der  t} = - \nabla p  + \eta \nabla^2 \vmes + \left(\zeta + \frac{\eta}{\dime}\right) \nabla \nabla \cdot \vmes + \nabla \cdot \tilde{\sigma},
\label{eq:mom}
\end{gather}

where $p$ is the pressure, $\rho$ is the mass density, $\vmes$ is the velocity,  $\dime=2,3$ is the dimension of the system, and $\eta$ and $\zeta$ are the shear and bulk viscosity, respectively. The term $\tilde{\sigma}$ in \eqref{eq:mom} corresponds to the random fluctuating stress tensor. The evolution of a concentration field, $\ca(\mathbf{r},t)$, of the species $\alpha$ (reactants or macromolecules) transported within the fluid, in a velocity field $\textbf{v}(\textbf{r},t)$, can be expressed in a general form in terms of the chemical potential $\mua$ as described by Ellero et al.\cite{Ellero2003}, and accounting for fluctuating compositional fluxes,\cite{Petsev2016} as 
\begin{equation}        
 \frac{\der \ca}{\der t}=-\ca(\boldsymbol{\nabla} \cdot \mathbf{v})+\boldsymbol{\nabla} \cdot \frac{T \ca}{\xi^{\alpha}} \boldsymbol{\nabla} \frac{\mua}{T} + \sa + \dfrac{\der\tilde{C}^{\alpha}}{\der t},  
\label{eq:compBalance} 
\end{equation}
where $\xi^{\alpha}$ is a transport friction coefficient, $T$ is the temperature, $\sa$ corresponds to source or sinks terms due to reactions, and ${\der\tilde{C}^{\alpha}}/{\der t}$ indicates the compositional fluctuations.  
Here, we follow the classical SPH kernel interpolation and the GENERIC formalism,\cite{Espanol2003,Ellero2003,Petsev2016} to discretize the momentum-balance equation \eqref{eq:mom} for the fluid particles, and the balance equation for the concentration \eqref{eq:compBalance} of the species $\alpha$. In \textcolor{black}{SDPD}, the discrete particles have a position ${\bf r}_i$ and velocity $\vmes_i$, and a volume $V_i$, such that $1/V_i = d_i = \sum_j W(r_{i j},h)$. Where $d_i$ is the number density of particles, $r_{ij} = |\mathbf{r}_i - \mathbf{r}_j|$
, and $W(r_{i j},h)$ is the normalized interpolant kernel with finite support $h$. In \textcolor{black}{SDPD}, a typical form of $W(r)$ can be written as \cite{Espanol2003}
\begin{equation}
W(r) = 
\begin{cases}{}
\frac{w_0}{h^\dime}\left(1+\frac{3r}{h}\right)\left(1-\frac{r}{h}\right)^3, \quad r/h<1.
\\
0, \quad r/h>1,
\end{cases}
\end{equation}
where $w_0=5/\pi$ or  $w_0=105/16\pi$ for two or three dimensions, respectively. 
We conveniently define the positive function $F_{ij} =- \nabla W(r_{i j},h)/r_{i j}$, to discretize \eqref{eq:mom}. Thus, the evolution equations for the particles position is ${\der {\bf{r}}_i}/{\der t} = \vmes_i$, whereas the momentum equation takes the form
\begin{align}
m_i\frac{\der \vmes_i}{\der t}  &= -\sum_j \left[ \frac{p_i}{d_i^2} + \frac{p_j}{d_j^2}\right] \fij \rbij \nonumber \\ 
& - \sum_j \left[\fa \vmes_{ij} +\fb (\vmes_{ij}\cdot\eij)\eij \right] \frac{\fij}{d_id_j} 
+ \dfrac{m_i\der\tilde{\vmes}_i}{\der t},
\label{eq:momsdpd}
\end{align}
where  $\eij = \rbij/|\rbij|$, $\vmes_{ij} = \vmes_i - \vmes_j$, $\fa$ and $\fb$ are friction coefficients related to the shear $\eta$ and bulk $\zeta$ viscosities of the fluid through $\fa={(\dime+2)\eta}/{\dime}-\zeta$ and $\fb = (\dime+2)(\zeta+{\eta}/{\dime})$, with $\dime$. The term $m_i\der \tilde{\vmes}_i/\der t$ in \eqref{eq:momsdpd} is given by\cite{Espanol2003}
\begin{align}
\dfrac{m_i\der \tilde{\vmes}_i}{\der t} = \sum_j \left(\Aij \der \wbbarij + \Bij \frac{1}{D}\text{tr}[ \der \wbij] \right) \cdot \frac{\eij}{\der t},
\end{align}
where $\wbbarij$ is a matrix of independent increments of a Wiener process for each pair $i,j$ of particles, and $\wbbarij$ is its traceless symmetric part, given by $\der \wbbarij = {1}/{2}\left[d\wbij+\der \wbij^T\right] - {\delta^{\alpha \beta}}/{\dime}\text{tr} [ \der \wbij].$
To satisfy the fluctuation-dissipation balance the amplitude of the thermal noises $\Aij$ and $\Bij$ are related to the friction coefficients $\fa$ and $\fb$ through
\begin{align}
A_{ij} &= \left[4\kbt \fa \frac{\fij}{\rhoi \rhoj} \right]^{1/2},
\\
B_{ij} &= \left[4\kbt\left(\fb -\fa\frac{\dime-2}{\dime}\right)\frac{\fij}{\rhoi \rhoj} \right]^{1/2},
\end{align}
where $\kbt$ is the thermal energy, and $\rhoi$ and $\rhoj$ are the densities of the particles $i$ and $j$. The pressure $p_i$ of the $i$-th particle is estimated using an equation of state of the form  
\begin{equation}
p_i = \frac{c^2\rho_0}{7}\left[ ({\rho_i}/{\rho_0})^{7}-1\right] +  \kappa\Pi_i +p_b,
\label{eq:eos}
\end{equation}
where $c$ is the artificial speed of sound on the fluid, and $\rho_0$ is the reference density. The term $c^2\rho_0/7$ correspond to the reference pressure of the system, where $c^2 = \partial p/ \partial \rho |_{\rho=\rho_0}$. The second term accounts for effects of the osmotic pressure,  $\Pi_i = C_i^{\alpha}\kbt$,\cite{VazquezQuesada2012sph} arising due to the presence of surrounding species $\alpha$, with concentration $C_i^{\alpha} = N_i^{\alpha}/V_i$. Throughout this document, we use superscripts $\alpha$ and $\beta$ to denote species or components of the system, while subscripts $i$ and $j$ refer to particle indices. In equation \eqref{eq:eos} $\kappa = {0,1}$ is a free parameter that allows to include osmotic pressure effects in the system.

Considering that the divergence of the velocity field in \textcolor{black}{SDPD} satisfies $(\boldsymbol{\nabla} \cdot \mathbf{v})_i = -\dot{d}_i/d_i$,\cite{Ellero2003} equation \eqref{eq:compBalance} can be expressed for a particle $i$ as
\begin{equation}        
    \frac{\der \ca_i}{\der t}= \left(\boldsymbol{\nabla} \cdot \frac{T \ca}{\xi^{\alpha}} \boldsymbol{\nabla} \frac{\mua}{T} + \sa + \dfrac{\der\tilde{C}^{\alpha}}{\der t} \right)_i.   
 \label{eq:constD}
\end{equation}
Assuming that the mass of the species transported within an \textcolor{black}{SDPD}-fluid element is significantly smaller than the mass of the \textcolor{black}{SDPD} particle, it is expected that $\der\tilde{C}_i^{\alpha} \ll \der{C}_i^{\alpha}$. As a consequence, it has been shown\cite{Li2015,Petsev2016,Drawert2019} that the fluctuation in the composition can be neglected ($\tilde{C}_i^{\alpha} \approx 0$). \textcolor{black}{This is appropriate in regimes where the number of molecules ($N_m$) per SDPD particle are sufficiently large and concentration fields can be regarded as smooth. In mesoscopic fluid domains with typical concentrations in the nM–mM range, the relative magnitude of concentration fluctuations scales as $1/\sqrt{N_m}$.\cite{van1992,gillespie2007} For example, for a concentration of 1 $\mu$M and an SDPD particle size of 100 nm, the number of molecules per particle $\sim 10^0$, leading to relative fluctuations close to ten percent. Increasing the concentration to 1 mM increases the number of molecules per particle to $\sim 10^4$, reducing relative fluctuations to about 0.4\%. Thus, in typical mesoscopic simulations with SDPD particles larger than a few tens of nanometers and concentrations above the micromolar range, concentration fluctuations can be safely neglected.}
    In general, for heterogeneous diffusion phenomena, the diffusion coefficients may depend on the concentration of $\alpha$, the presence of other species, time, and the direction of flow. However, for simplicity, we neglect inter-diffusivities among different species, assuming that diffusivities are independent of each other. \textcolor{black}{This approximation is valid when solute–solute interactions are weak and transport is governed primarily by molecular size or hydrodynamic interactions with the solvent. In crowded or strongly interacting systems (e.g., ionic mixtures, gels, or cytoplasmic environments), cross-diffusion terms become important and cross-diffusion descriptions, as introduced by Thieulot et al.\cite{Thieulot2005} can be consistently incorporated into our SDPD framework.}

Consequently, the discrete \textcolor{black}{SDPD} formulation of \eqref{eq:constD} can be expressed as
\begin{equation}
\frac{\der {C}_i^{\alpha}}{\der t}  = \sum_j d_{\text{eq}}D_{ij}^{\alpha}  C_{ij}^{\alpha} \frac{\fij}{d_id_j} + S_i^{\alpha},
\label{eq:compBalanceSDPD}
\end{equation}
where $D_{ij}^{\alpha} = {4D_i^{\alpha}D_j^{\alpha}}/{(D_i^{\alpha}+D_j^{\alpha})}$, and $C_{ij} = C_i-C_j$. Here, $d_{\text{eq}}$ is the equilibrium particle number density, $D_i^{\alpha}$ and $D_j^{\alpha}$ corresponds to the diffusion of the ${\alpha}$-th species on particles $i$ and $j$. Notice that if the diffusion of the species is the same among particles, $D_{ij}^{\alpha} = 2D^{\alpha}$.

We adopt a system of dimensionless (reduced) units by defining characteristic scales for length ($\ulen$), mass ($\umass$), and energy ($\uener$). This leads to a natural unit of time given by $\utime = \ulen \sqrt{\umass/\uener}$. All physical quantities are expressed in terms of these reference units. For instance, a domain length $L_x = 20$ corresponds to a physical size of $20\,\ulen$, while an interparticle spacing of $dx = 0.2$ implies a resolution of $0.2\,\ulen$. Transport and reaction parameters are similarly nondimensionalized: diffusion coefficients are expressed in units of $\ulen^2/\utime$, reaction rates in $1/\utime$, and velocities in $\ulen/\utime$. This consistent framework facilitates comparison across all test cases.

To streamline the presentation of our results, we express several key dimensionless numbers. The Reynolds number, $\mathrm{Re} = \rho u \ulen / \eta = u \ulen / \nu$, characterizing the ratio of inertial to viscous forces, where $\rho$ is fluid density, $u$ is a characteristic velocity, $\eta$ is the dynamic viscosity, and $\nu$ the kinematic viscosity. The Péclet number, $\mathrm{Pe} = u \ulen / D$, which quantifies the relative strength of advection versus diffusion, and the Schmidt number, $\mathrm{Sc} = \nu / D$, relates momentum transport to mass transport. Finally, the Damk\"ohler number, $\mathrm{Da} = \ulen^2 / (D \tau_{\text{r}})$, compares the timescales of diffusion and chemical reaction, with $\tau_{\text{r}}$ denoting the characteristic reaction time.
\section{Results}\label{sec:results}

\subsection{Diffusion-dominated validation}\label{subsec:benchmarks-Dif}

Without loss of generality, we present our simulation results in one- and two-dimensional domains to validate the implemented ADR-\textcolor{black}{SDPD} scheme for diffusion-dominated systems. It is important to note that in equation \eqref{eq:compBalanceSDPD}, the diffusion coefficient $D^{\alpha}$ for each species in the system is defined as an input parameter for the \textcolor{black}{SDPD} discretization. However, as observed from the momentum balance in \eqref{eq:momsdpd}, \textcolor{black}{SDPD} particles also exhibit inherent diffusion due to their thermal motion,\cite{Litvinov2009} which contributes to the overall species transport. Thus, we define an effective diffusion coefficient, $D_{\text{eff}}^{\alpha}$, to account for the combined effect of deterministic ($D_{\text{det}}^{\alpha} = D^{\alpha}$) and stochastic ($D_{\text{stoch}}^{\alpha}$) transport. We first validate the correctness of our implementation to model the deterministic contribution by setting $\kbt = 0$, in one-dimensional domains. Then we assess the effects of both deterministic and stochastic contributions on $D_{\text{eff}}^{\alpha}$ for $\kbt > 0$. 

\subsubsection*{Validation of the Deterministic Contribution}

For this validation, we consider a periodic channel with Dirichlet and Neumann wall boundary conditions using sinusoidal initial conditions, and compare the temporal evolution of the concentration profiles \(C(x,t)\) with their corresponding analytical solutions. To facilitate consistent interpretation across all benchmark tests, we adopt fixed characteristic scales: $\ulen = 1$ and $\utime = 1$. All reported quantities are expressed in reduced units, unless otherwise specified.

\begin{figure}[!hbtp]
\centering
\includegraphics[width=.48\textwidth]{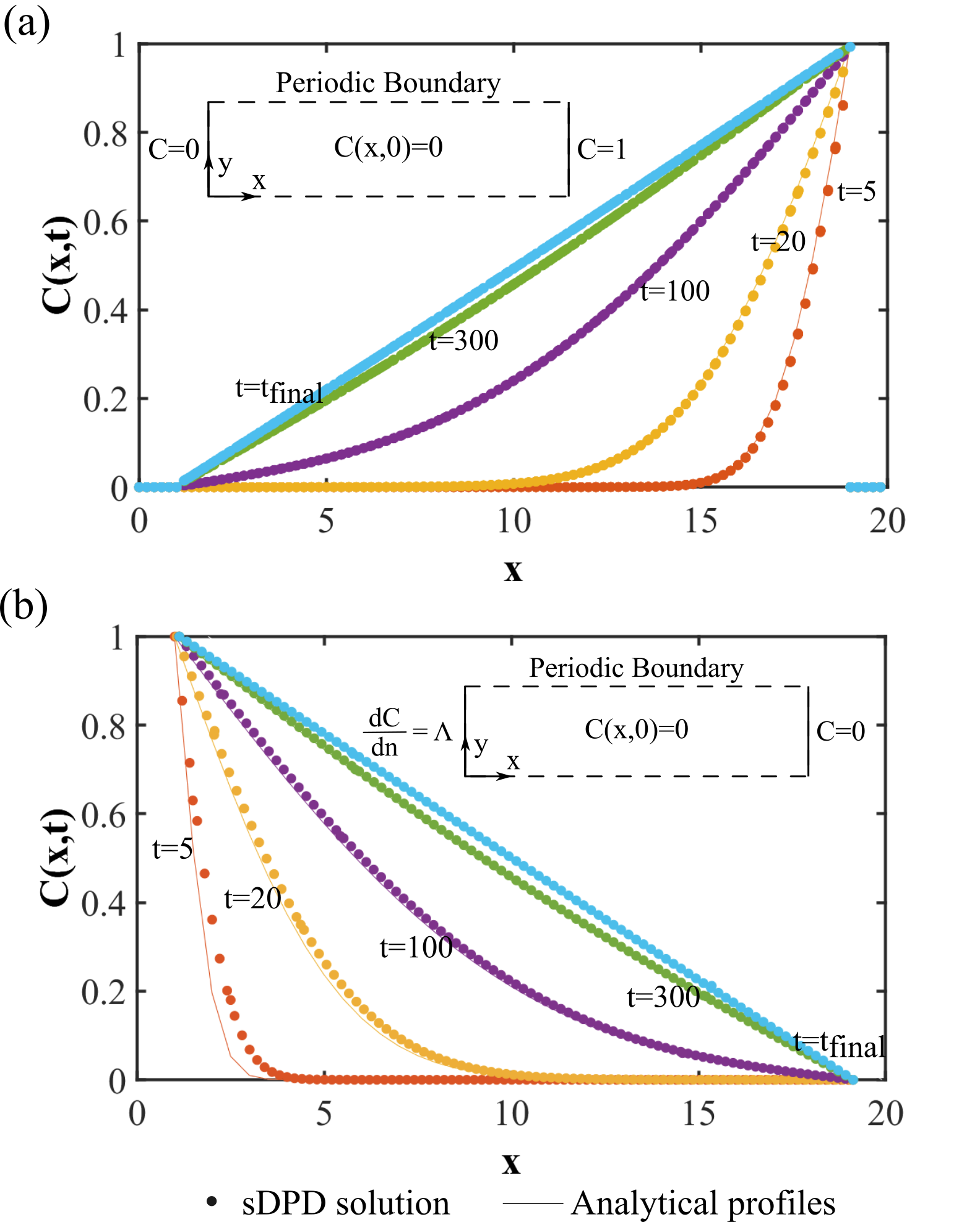}
\caption{Temporal evolution of the concentration profile $C(x,t)$ at $t = [5, 20, 100, 300, 500]\,\tau$ for (a) Dirichlet and (b) Neumann boundary condition tests. Numerical results are compared against analytical solutions (see Appendix \ref{app:Difftest1D}). Insets depict the respective channel configurations.  Simulation parameters are listed in Table \ref{tab:sdpd-params}.}
\label{fig:RD_Diff_tests_sdpd_1D}
\end{figure}

The simulations are performed in a domain of size $L_x \times L_y = 20\ulen \times 10\ulen$, with periodic boundaries in the $y$-direction and solid walls along $x$, enforcing no-slip velocity and no-flux concentration conditions. In Fig.\ref{fig:RD_Diff_tests_sdpd_1D}, we present the concentration profiles \(C(x,t)\) simulated at five different times (\(t = [5, 20, 100, 300, 500]\,\tau\)), for Dirichlet (Fig.\ref{fig:RD_Diff_tests_sdpd_1D}a) and Neumann (Fig.\ref{fig:RD_Diff_tests_sdpd_1D}b) boundary conditions.

In general, from Fig.\ref{fig:RD_Diff_tests_sdpd_1D}, we can observe that our \textcolor{black}{SDPD} implementation consistently reproduce the corresponding analytical solutions\cite{crank1975mathematics} (see Appendix \ref{app:Difftest1D}), for each of the BCs implemented.  The \textcolor{black}{SDPD} parameters used for the simulations in this study include $\kbt = 0$, $\rho = 1.0$, $\eta = 3.0$, and $c_0 = 50$, as summarized in Table \ref{tab:sdpd-params} (Appendix \ref{app:SItables}).

\subsubsection*{Effective Diffusion}\label{subsubsec:dif2dvalidation}

\begin{figure}[!hbtp]
\centering
\includegraphics[width=.5\textwidth]{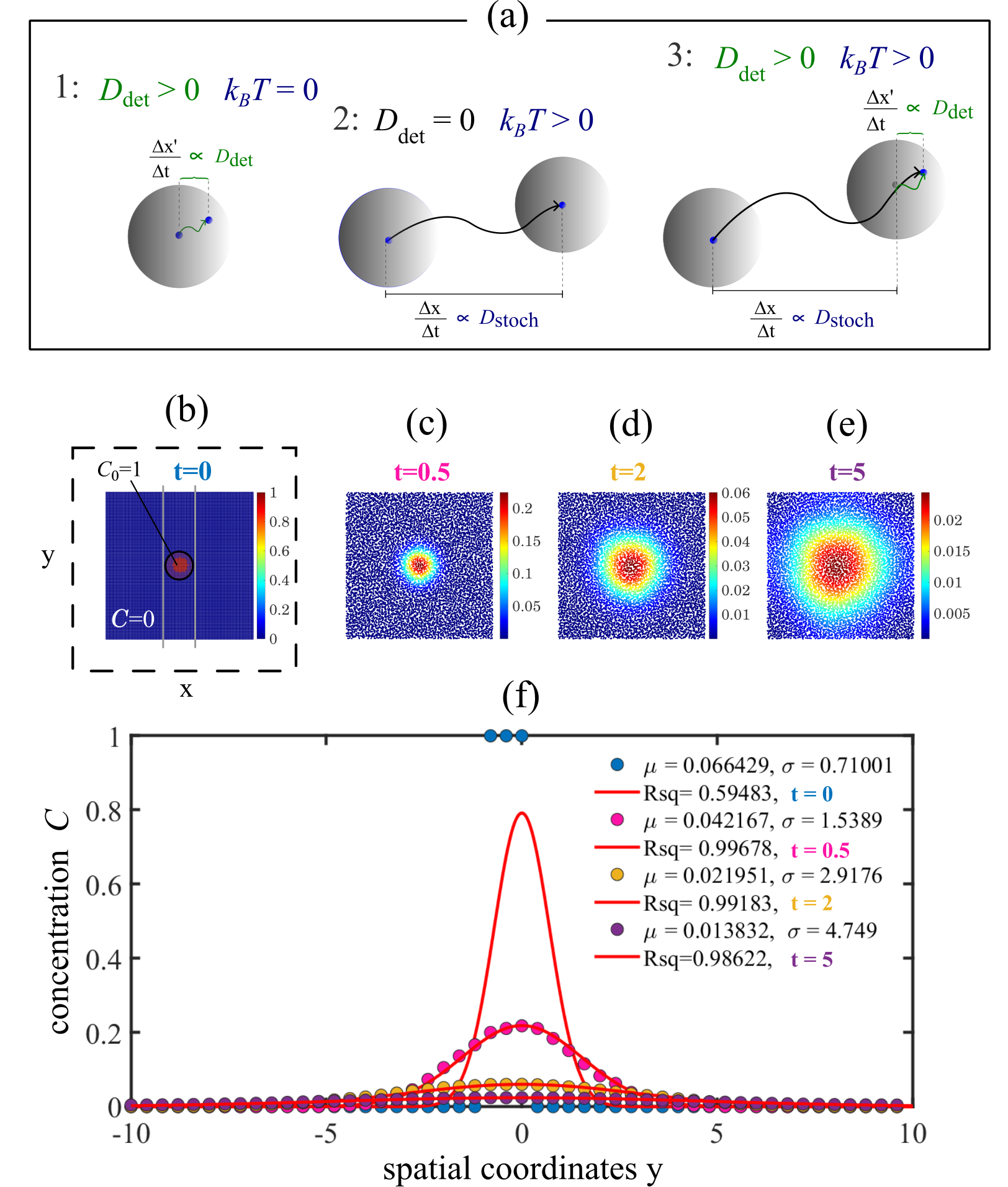}
\caption{Computational validation for 2D diffusion. Panel (a) illustrates the three distinct simulation schemes, each with a large gray sphere representing an \textcolor{black}{SDPD} particle and a smaller blue sphere inside to symbolize a molecule of the chemical substance. These schemes investigate deterministic ($D_{\mathrm{det}}$) and stochastic ($D_{\mathrm{stoch}}$) diffusion, with temperature \(\kbt\) as a variable.  Subfigure (f) displays Gaussian profiles with estimated parameters --mean ($\mu$) and standard deviation ($\sigma$)-- along with the corresponding $R^2$ values, captured at times $t$=0, $t$=0.5, $t$=2 and $t$=5. On top of (f), the scatter plots (b)-(e) represent the concentration distribution at these time points. Subfigure (b) also highlights the periodic boundary conditions applied to the 2D computational domain. \(\kbt = 1\), \(D^{\alpha} = 2\).}
\label{fig:RD_Diff_tests_sdpd_sketches2D}
\end{figure}

We now investigate whether the effect of deterministic and stochastic contributions on $D_{\text{eff}}^{\alpha}$ is additive in nature, or whether it can give rise to non-trivial diffusional behavior. To explore this, we consider three distinct simulation schemes isolating the different diffusion mechanisms. In scheme 1, we set $\kbt = 0$, to represent the conditions where $D_{\text{eff}}^{\alpha} \to D_{\text{det}}^{\alpha}$. In scheme 2, we set $D_{\text{det}}^{\alpha} = 0$, and varied the thermal energy of the system $\kbt > 0$, such that $D_{\text{eff}}^{\alpha} \to D_{\text{stoch}}^{\alpha}$. Finally, in scheme 3, we combine both effects, incorporating non-zero input diffusion and non-zero temperatures. These setups are schematically represented in Fig. \ref{fig:RD_Diff_tests_sdpd_sketches2D}a, providing a framework for comparing the contributions of deterministic and stochastic diffusion to the effective diffusion coefficients.

The simulations were performed in 2D square domains of side length $l_{\text{box}} = 20\ulen$. Each domain contained a single chemical species, initially distributed as a droplet of radius $r$ centered in the domain. The initial concentration was uniform within the droplet, with a value $C = C_0$, as shown in Fig. \ref{fig:RD_Diff_tests_sdpd_sketches2D}b. The evolution of the concentration profile $C(x, t)$ was tracked for $D^\alpha = 2$ and $\kbt = 1$, with snapshots shown in Fig. \ref{fig:RD_Diff_tests_sdpd_sketches2D}c-e at times $0.5\tau$, $2\tau$, and $5\tau$, respectively. At each time, the concentration profile could be well approximated by a Gaussian distribution of the form:
\begin{equation}
C(x,t) = C_0 \exp\left(-\frac{(x - x_0)^2}{2\sigma^2}\right),
    \label{eq_RD_GaussianEq}
\end{equation}
where $C_0$ is the peak concentration, $x_0$ is the center position, and the variance $\sigma^2$ satisfies $\sigma^2 = 2 D_{\text{eff}}^{\alpha} t$. Accordingly, the effective diffusion coefficient $D_{\text{eff}}^{\alpha}$ was estimated by fitting the concentration profiles to Eq. \eqref{eq_RD_GaussianEq}. An example of this estimation is shown in Fig. \ref{fig:RD_Diff_tests_sdpd_sketches2D}f for $\kbt = 1$ and $D_{\text{det}}^{\alpha} = 2$, at times $t = [0, 0.5, 2, 5]\,\tau$.

To explore the interplay between deterministic and stochastic diffusion, we simulated four species with different diffusion coefficients. For slower-diffusing species, simulations were run up to $t = 10\tau$, while shorter durations sufficed for faster diffusion. In scheme 1, we varied the deterministic diffusivity $D_{\text{det}}^{\alpha} = 2,\ 1,\ 0.5,\ \text{and } 0.25\ l^2/\tau$. In scheme 2, with $D_{\text{det}}^{\alpha} = 0$, we modulated thermal noise by setting $k_BT = 1,\ 3,\ \text{and } 5$. In scheme 3, we combined the non-zero deterministic diffusion from scheme 1 with the thermal noise from scheme 2. All simulations were conducted with a fixed kinematic viscosity $\nu = 3.0$, yielding Schmidt numbers $\mathrm{Sc} = \nu / D^{\alpha}$ of 1.5, 3.0, 6.0, 12. 

\begin{figure}[!htbp]
\centering
\includegraphics[width=.48\textwidth]{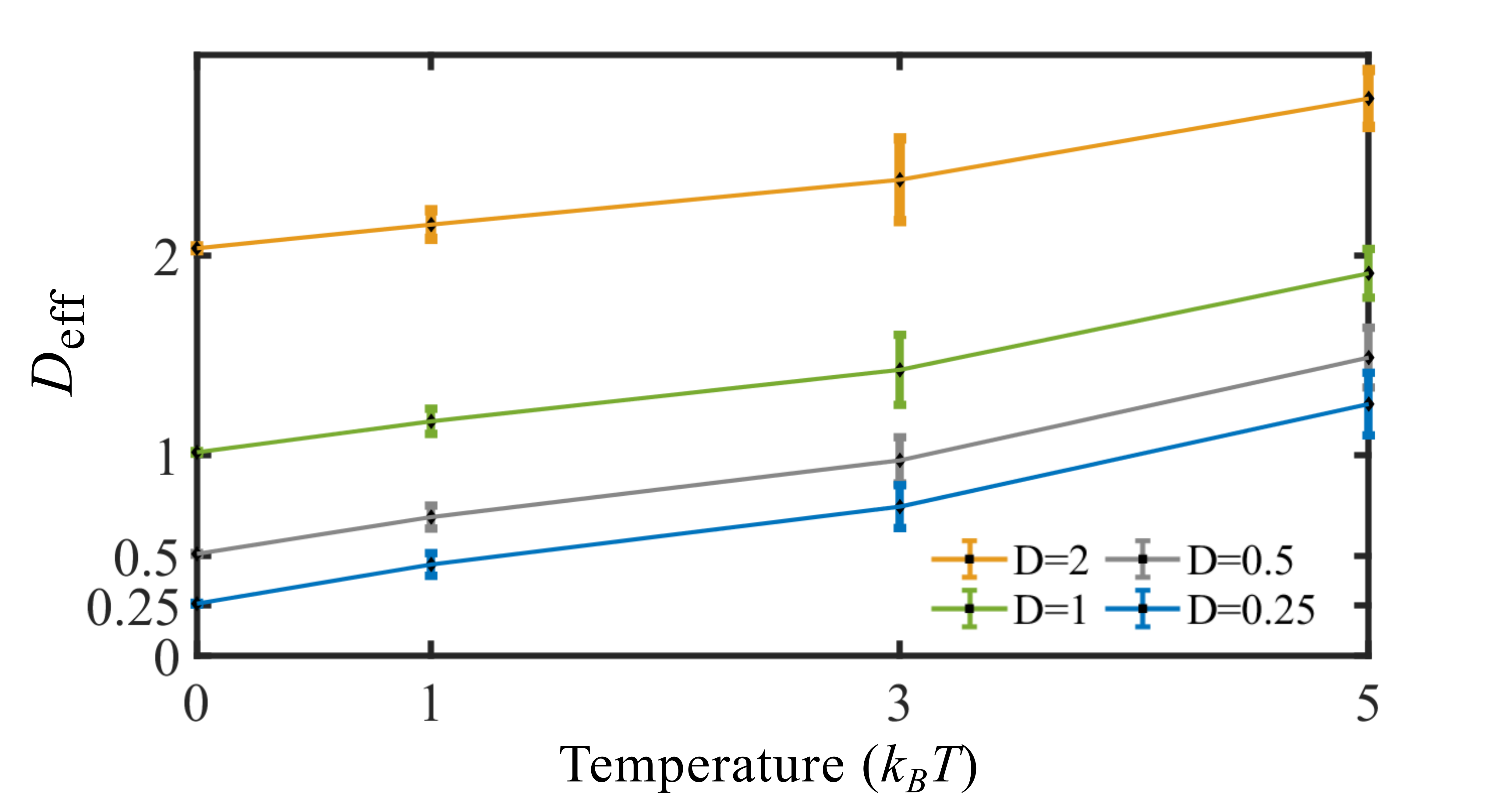}
\caption{Estimated effective diffusion coefficient ($D_{\text{eff}}$) for different input values of $D$, plotted across a range of temperatures ($\kbt = 0, 1, 3, 5$). The results illustrate the transition from stochastic to deterministic diffusion, with $D_{\text{eff}}$ approaching the deterministic limit as $\kbt \to 0$. Line colors indicate input $D$ values.}
\label{fig:RD_Validation_Deff_3Sets}
\end{figure}

In Fig. \ref{fig:RD_Validation_Deff_3Sets}, we summarize the variation of $D_{\text{eff}}^{\alpha}$ across the different schemes investigated. In general, $D_{\text{eff}}^{\alpha}$ exhibit a linear dependency with the temperature for a fixed deterministic diffusion. Our analysis shows that the deterministic $D_{\text{det}}^{\alpha}$ and stochastic  $D_{\text{stoch}}^{\alpha}$ diffusion, up to a good approximation, contribute additively to the effective diffusion (see Appendix \ref{app:Difftestsexpanded}, Table \ref{tab:RD_difftest_Temp_1_3_5} for details), leading to
\begin{equation}
D_{\text{eff}}^{\alpha} \approx D_{\text{det}}^{\alpha} + D_{\text{stoch}}|_{\kbt}.
\label{eq:Diff_Sum_det_stoch}
\end{equation}
Under deterministic-only conditions ($\kbt$=0), the additive approximation for effective diffusivity, \eqref{eq:Diff_Sum_det_stoch}, holds with excellent accuracy, yielding relative errors below 2$\%$ for input \( D^\alpha = 2\) and \( D^\alpha = 1 \). However, when thermal fluctuations are introduced ($\kbt$=1 and 5), relative errors increase to 6–9$\%$ (see Appendix \ref{app:Difftestsexpanded}, Table \ref{tab:RD_difftest_Temp_1_3_5}).

\subsection{Reaction-Dominated Systems}\label{subsec:ReacValidation}

We validate the implemented \textcolor{black}{SDPD} scheme by simulating three reaction-dominated benchmarks, each governed by distinct kinetics: (a) a simple equilibrium reaction ($2\mathrm{SO_2} + \mathrm{O_2} \rightleftharpoons 2\mathrm{SO_3}$); (b) a reaction network with parallel and series steps\cite{UMichiganPFR} ($\mathrm{A} + 2\mathrm{B} \rightarrow \mathrm{C}$, followed by $3\mathrm{C} + 2\mathrm{A} \rightarrow \mathrm{D}$); and (c) an enzymatic substrate conversion described by Michaelis–Menten kinetics, involving the formation of an intermediate enzyme–substrate complex ($\mathrm{E} + \mathrm{S} \rightleftharpoons \mathrm{ES} \rightarrow \mathrm{E} + \mathrm{P}$).

For each system, the source terms $S^{\alpha}$ for species $\alpha$ are defined according to the compositional balance in Equation \eqref{eq:compBalanceSDPD}. A detailed description of the kinetic formulations is provided in Appendix \ref{app:kinetics}. Simulations are performed in fully periodic square domains of side length $l_{\text{box}} = 10\ulen$, using a uniform fluid particle distribution.

To enable comparison across systems with different reaction rates, we present the results in terms of the dimensionless time $t/\tau_{\text{r}}$, where the characteristic reaction time is defined as $\tau_{\text{r}} = 1 / k_{\alpha}^{\text{max}}$, with $k_{\alpha}^{\text{max}}$ denoting the largest reaction rate constant in each system. To characterize the competition between diffusion and reaction over the full domain, we define the Damköhler number as $\mathrm{Da} = {l_{\text{box}}^2}/{D\, \tau_{\text{r}}}$. For all benchmark cases, we find $\mathrm{Da} \gg 1$, indicating a reaction-dominated regime on the scale of the domain. Notation and units are consistent with those defined in Appendix \ref{app:SItables}.
  
In Fig. \ref{fig:RD_Kinetictests_abc}, we summarize the compositional evolution for the different kinetics observed in our \textcolor{black}{SDPD} simulations and compare them with the corresponding ODE solution estimated in MATLAB using \texttt{ode45}. Fig. \ref{fig:RD_Kinetictests_abc}a, for an equilibrium kinetics, shows that when the initial concentration of the species is at equilibrium, the system remains in a steady state, as expected. In contrast, reducing the initial concentration of $\mathrm{SO_3}$ leads the system to evolve toward a new equilibrium that satisfies the theoretical equilibrium constant $K_{\mathrm{c}} = 4.3$, in agreement with predictions from classical thermodynamics.\cite{stockmayer1944} Fig. \ref{fig:RD_Kinetictests_abc}b, modelling a more complex kinetics, evidence that the time evolution of the species concentrations obtained from \textcolor{black}{SDPD} simulations closely matches the corresponding solutions of the ordinary differential equations (ODEs), demonstrating excellent quantitative agreement. Finally, Fig. \ref{fig:RD_Kinetictests_abc}c, depicting the response of an enzymatic kinetics, shows that the concentration profiles follow theoretical expectations for Michaelis–Menten dynamics,\cite{Qian2021} further validating our \textcolor{black}{SDPD} implementation to accurately capture a range of kinetics, including nonlinear biochemical processes.

\begin{figure}[!htbp]
\centering
\includegraphics[width=.41\textwidth]{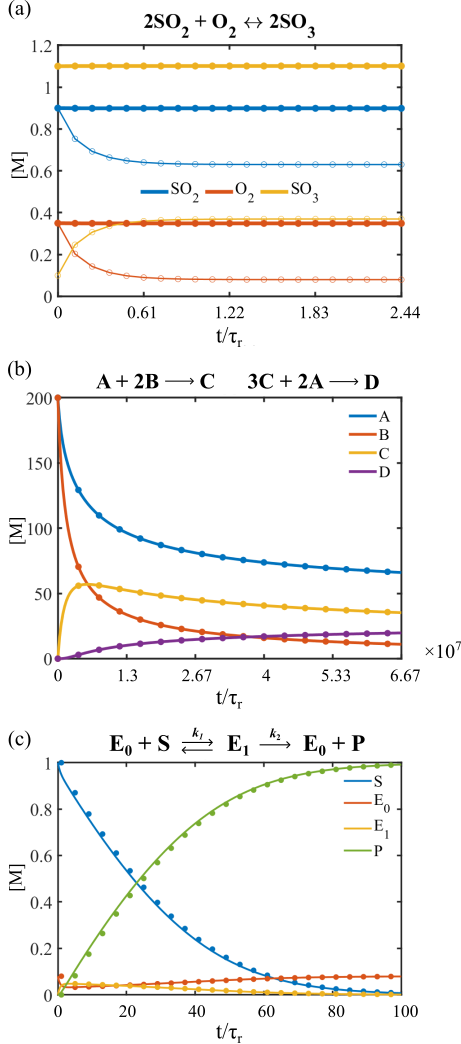}
\caption{Mean concentration profiles $C^{\alpha}$ of each chemical species $\alpha$ over dimensionless time ($t^* = t/\tau_{\mathrm{r}}$), with concentrations in Molar units. Solid lines show ODE solutions; circles indicate \textcolor{black}{SDPD} results. (a) Sulfur trioxide formation: initial equilibrium state (thick lines, filled circles) and evolution toward a new equilibrium after perturbation (thin lines, open circles). (b) Temporal evolution of species in the complex reaction system originally designed for a plug-flow reactor (PFR). (c) Enzyme-substrate dynamics according to the Michaelis–Menten model.}
\label{fig:RD_Kinetictests_abc}
\end{figure}

\subsection{Coupled Advection-Diffusion}  

\begin{figure}[!htbp]
\centering
\includegraphics[width=.5\textwidth]{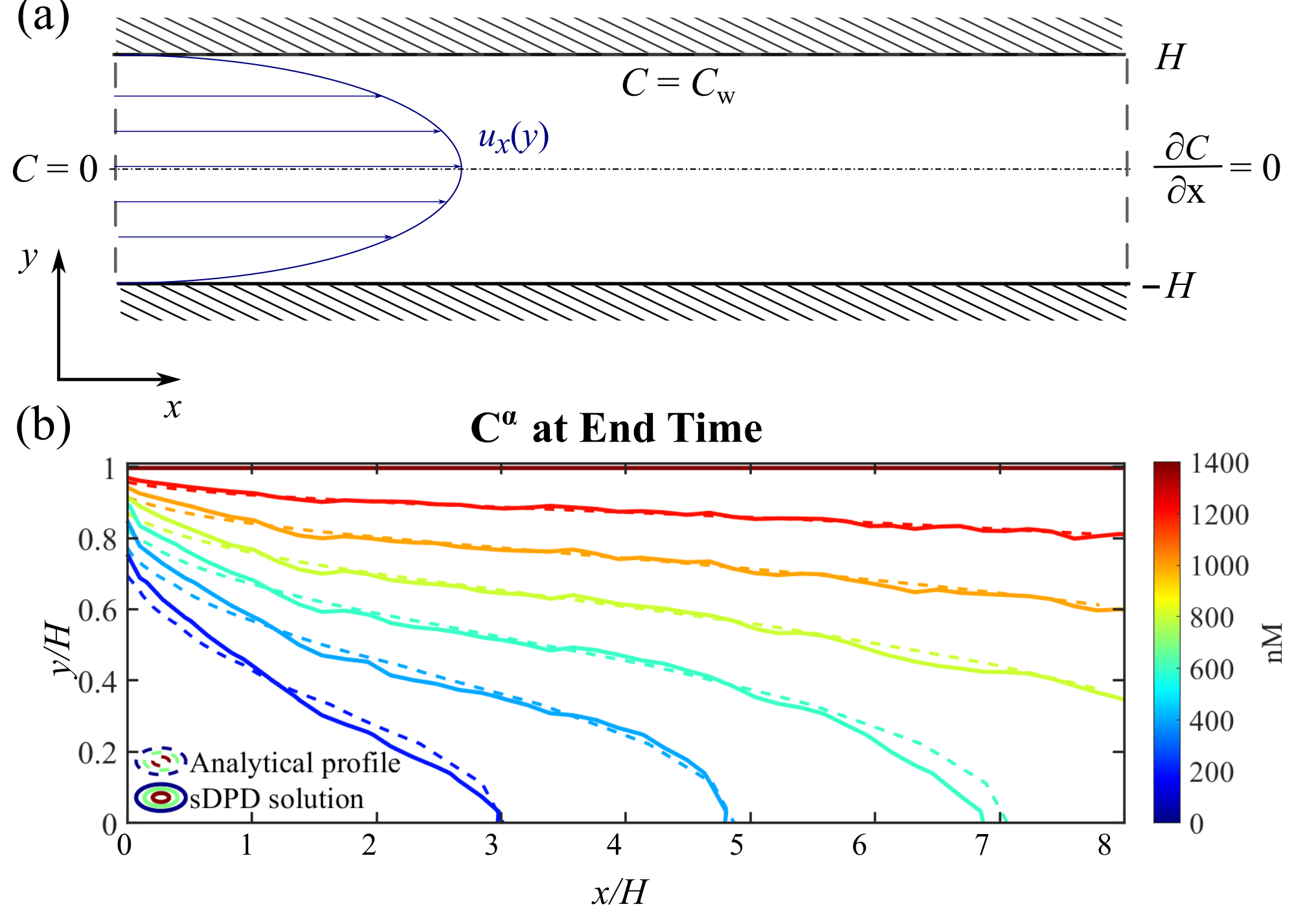}
\caption{Top: sketch of the channel geometry for the Poiseuille flow configuration, depicting the characteristic parabolic velocity profile and the boundary conditions for the concentration. Bottom: comparison of analytical concentration contours and simulation results for $C^\alpha$. Simulation parameters: \(u_0 = 0.24\), \(D = 0.22 \ulen^2/\tau\), with $L_y/2=H$ and $H=5$. $L_x=40$.}
\label{fig:RD_2d_Poiseuille}
\end{figure}

In the preceding section, we explored spatial diffusion and interaction of chemical species under static conditions. Now, we validate our framework considering the transient development of a concentration profile under flow conditions. This scenario is governed by the steady-state advection-diffusion equation, which captures the interplay between molecular diffusion and convective transport. As illustrated in Fig. \ref{fig:RD_2d_Poiseuille}a, the boundary conditions for the concentration field on the channel walls are as follows: \(C(\pm H) = C_\mathrm{w}\), where \(C_\mathrm{w} = 1400\) nM represents a prescribed equilibrium concentration at the channel walls.\cite{ratto2021patient} At the inlet, \(C = 0\), indicating that the species is introduced downstream from a region of negligible concentration. At the outlet, a zero-gradient condition, \(\der C / \der x = 0\), ensures that the concentration profile remains stable and unaffected by further downstream flow.

In Fig.\ref{fig:RD_2d_Poiseuille}b, we present the concentration profiles obtained from the \textcolor{black}{SDPD} simulation for half of the channel and compare them with the analytical solution derived by Kuzmin et al.,\cite{kuzmin2013lattice} applying a symmetry boundary condition (see Appendix \ref{app:advdiff}, Eq.\eqref{eq_RD_ADE_analyticsol}). The test is performed under moderate advection conditions, characterized by $u_0 = 0.24$, $H = 5\,\ulen$, $D = 0.22\,\ulen^2/\utime$, and $\nu = 3.0\,\ulen^2/\utime$. These parameters correspond to the following dimensionless numbers: Péclet number $\mathrm{Pe} \approx 11$, Reynolds number $\mathrm{Re} = 0.4$, and Schmidt number $\mathrm{Sc} \approx 14$. The good agreement between simulation and analytical results confirms that our numerical method accurately captures the benchmark case.

\subsection{Turing-Pattern Formation}\label{subsec:TuringPattern}

Finally, we explore a complex system where Turing-like patterns emerge, and investigate the effect of thermal fluctuations on the stability of such patterns. This setting couples nonlinear reaction kinetics with diffusion to study the emergence of spatially organized patterns in biologically inspired system. The model is adapted from the study of activator (${Ac}$) – inhibitor (${In}$) dynamics,\cite{cooper2018ancient} which explores the formation of self-organized chemical patterns driven by local excitation and long-range inhibition. A complete description of the governing equations, source terms, and initial and boundary conditions is provided in Appendix \ref{app:ReacDiffEquations}. 

We simulated a square domain of side length $L_s = 75\ulen$, initialized with uniform particle spacing $dx = 0.2\ulen$. The system was evolved over a physical simulation time $t = 1500$, using a time step $\Delta \tau = 0.001$. To facilitate nondimensional analysis, time is rescaled by the diffusive timescale of the inhibitor species, defined as $t^* = L_s^2 / D^{In}$. The corresponding dimensionless time is $\mathcal{T} = t / t^*$, and all results are presented in terms of this rescaled time variable.

\begin{figure}[!h]
\centering
\includegraphics[width=.48\textwidth]{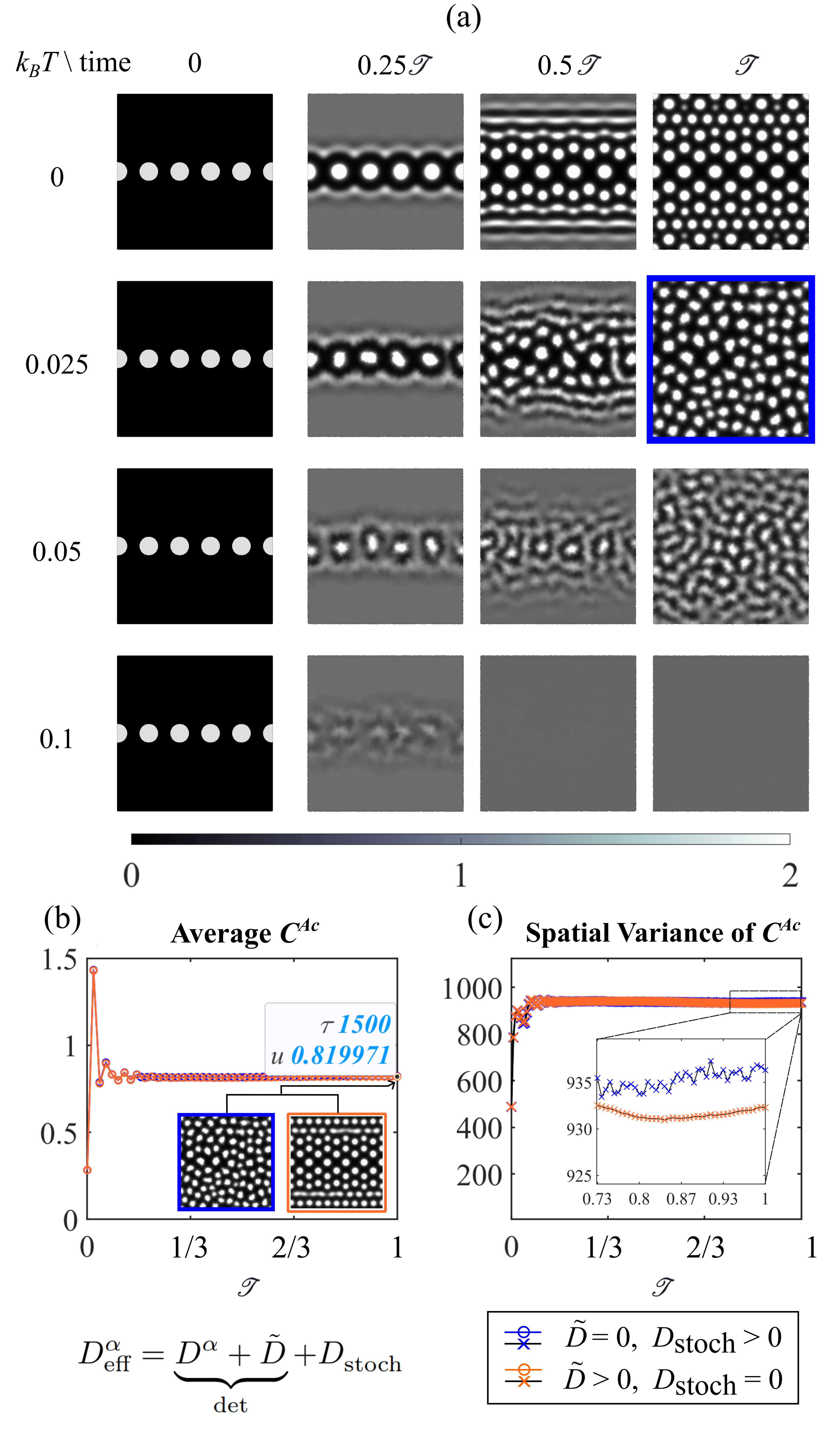}
\caption{\small{Pattern formation in the Reaction-Diffusion (RD) system for the activator $C^{Ac}$.
(a) Representative pattern evolution at four timepoints, starting from initial activator spots. Rows correspond to increasing values of thermal noise $\kbt$, illustrating how higher thermal noise suppresses pattern formation.
(b) Time evolution of the spatial average of $C^{{Ac}}$ and final pattern insets for two cases: the hybrid stochastic–deterministic reference (blue) with $\kbt = 0.025$ and $D^{{Ac}}_{\text{det}} = 0.02$, and an athermal simulation (orange) with $\kbt = 0$ and $D^{{Ac}}_{\text{eff}} = D^{{Ac}}_{\text{det}} + \tilde{D}_{\text{det}} = 0.0263\, \ulen^2/\tau$. Both match in average and variance but exhibit different pattern onset times.
(c) Spatial variance of $C^{Ac}$ over time, showing convergence to a comparable quasi-steady state in both simulations. A zoom-in highlights reduced temporal fluctuations in the athermal case, consistent with the absence of intrinsic noise.
}
}
\label{fig:RD_TuringbenchmarkAB}
\end{figure}

The simulations are initialized with six equally spaced activator spots ($C^{Ac}$) placed along the horizontal midline of the domain (see Fig. \ref{fig:RD_TuringbenchmarkAB}a, first column),  while the inhibitor concentration $C^{In}$ is initially uniform. The parameters used match those reported in,\cite{Miura2004} specifically selected to promote pattern formation via a Turing mechanism. The dimensionless Damk\"ohler numbers computed from the system parameters, $\mathrm{Da}^{Ac} = k^{Ac}_{\text{deg}} {(L_s)}^2 / D^{Ac} = 8437.5$ and $\mathrm{Da}^{In} = k^{In}_{\text{deg}} {(L_s)}^2 / D^{In} = 750$, indicate a strongly reaction-dominated regime for both species. This separation of timescales is consistent with conditions that promote Turing instabilities in reaction–diffusion systems. \cite{turing1952, murray2000pattern} A classical requirement for such instabilities is that the inhibitor diffuses significantly faster than the activator, while its reaction kinetics are \emph{not overly dominant}, ensuring that local activation and long-range inhibition can emerge. Specifically, this corresponds to $D^{In} \gg D^{Ac}$ and $k^{In}_{\text{deg}} \lesssim k^{Ac}_{\text{deg}}$. \cite [see Ch.~2]{murray2000pattern}. In our case, while the diffusion disparity is well satisfied with $D^{In}/D^{Ac} \approx 30$, the degradation rate ratio $k^{In}_{\text{deg}}/k^{Ac}_{\text{deg}} \approx 2.67$ suggests that inhibition is not slower, but still operates on a sufficiently comparable timescale to allow pattern formation.
In Fig. \ref{fig:RD_TuringbenchmarkAB}a, we present the evolution of the activator concentration for different thermal fluctuations levels, $k_BT = \{0.0 , 0.025, 0.05, 0.1\}$. Simulations in athermal conditions ($\kbt=0$) successfully reproduced activator patterns formation that are consistent with previously reported results.\cite{cooper2018ancient} As $k_B T$ increases, the effective diffusivity $D_{\text{eff}}^{\alpha}$ of both species also increases due to the stochastic contribution $D_{\text{stoch}}$, leading from weak disruptions of the patterns at $\kbt=0.025$, to their gradual smearing as we reach $\kbt=0.1$. As expected, the enhanced transport induced by $D_{\text{stoch}}$, alters the Damk\"ohler number of the species and the delicate balance between activator and inhibitor diffusion-reaction. 

Interestingly, at modest thermal fluctuations ($\kbt=0.025$), the system seems to remain near the critical conditions for Turing-pattern formation (see Fig. \ref{fig:RD_TuringbenchmarkAB}a). However, those patterns exhibit noticeable perturbations on their symmetry and sharpness. To understand the role of the thermal fluctuations over the pattern formation, we verify if the additivity of the diffusion contributions \eqref{eq:Diff_Sum_det_stoch} remains valid for complex diffusion-reaction conditions. To this end, in \ref{fig:RD_TuringbenchmarkAB}b–c, we compare the results obtained for a system with $\kbt=0.025$, with respect to an equivalent \textit{purely deterministic} setup ($\kbt=0$). For the stochastic-deterministic case, we have $D_{\text{eff}}^{\alpha} \approx D_{\text{det}}^{\alpha} + D_{\text{stoch}}|_{0.025}$. Whereas, for the purely deterministic, we set an effective diffusion $D_{\text{eff}}^{\alpha} = D_{\text{det}}^{\alpha} + \tilde{D}_{\text{det}}$, where $\tilde{D}_{\text{det}} = D_{\text{stoch}}|_{0.025}$ is a deterministic diffusion with the same magnitude of the stochastic contribution at $\kbt=0.025$. We estimate $D_{\text{stoch}}|_{0.025} = 0.0063\, l^2/\tau$ from the mean squared displacement (MSD) analysis (see Appendix \ref{appB:MSD}) of the particles in a simulation at $\kbt=0.025$. Notice that the term $\tilde{D}_{\text{det}}$ is added to both species, activator ($D_{\text{det}}^{Ac} = 0.02\, l^2/\tau$) and inhibitor ($D_{\text{det}}^{In} = 0.6\, l^2/\tau$), leading to total input diffusivities ($D^\alpha_{\text{eff}}$) of 0.02625 and 0.60625, respectively.

In Fig.\ref{fig:RD_TuringbenchmarkAB}b–c, we summarize the evolution of the average and spatial variance of activator species (${Ac}$), for the stochastic–deterministic reference simulation (blue) and the purely deterministic case (orange). 

In general, we observe that both schemes reproduce a comparable temporal average. However, although both simulations converge to similar mean and variance values at late times, the purely deterministic case exhibits a smoother evolution with reduced fluctuations in the spatial variance (Fig. \ref{fig:RD_TuringbenchmarkAB}c). This behavior reflects the absence of intrinsic thermal noise, which dampens spatiotemporal variability in the emerging patterns. Thus, even though both systems yield an equivalent effective diffusivity $D_{\text{eff}}^{\alpha}$ and match global statistical metrics, the perturbations introduced by thermal fluctuations can significantly disrupt the morphogen distribution.

These differences become more evident when comparing the final patterns: purely deterministic systems (orange) generate underdeveloped spatial structures relative to the reference stochastic–deterministic simulations (blue). This suggests that in the presence of thermal fluctuations, the effective diffusivity may be greater than the sum $D_{\text{det}} + \tilde{D}_{\text{det}}$, likely due to nonlinear interactions between stochastic fluctuations and reaction kinetics that are not captured by the additive model.

Overall, our results indicate that thermal fluctuations can impact patterning beyond what diffusivity alone can explain.

\section{Conclusions}\label{sec:discussion}
The \textcolor{black}{SDPD} framework presented in this work offers a flexible and thermodynamically consistent tool for modeling reaction–diffusion systems at mesoscopic scales. A key strength of the method lies in its ability to independently calibrate input diffusivities through particle-level interactions, enabling precise control over species-specific transport. This is particularly useful for investigating systems where stochastic and deterministic transport coexist, as it allows isolating the role of thermal fluctuations from externally imposed diffusion. Leveraging this capability, we proposed a simple strategy to match stochastic and athermal transport regimes by estimating the thermal diffusivity contribution via mean squared displacement (MSD) analysis.

We validated the method across a range of canonical problems--from pure diffusion in 1D and 2D, to reaction-only kinetics, and coupled advection–diffusion–reaction scenarios. These benchmarks demonstrate the robustness of the implementation and its ability to capture relevant dynamical features. In particular, our analysis of Turing pattern formation highlights that thermal noise influences not only effective transport but also the timing and structure of emerging spatial patterns. While the additive approximation $D_{\text{eff}}^{\alpha} \approx D_{\text{det}}^{\alpha} + D_{\text{stoch}}$ is valid in terms of average system behavior, similar global diffusivities do not necessarily imply equivalent spatial outcomes, as fluctuations may interact nonlinearly with reaction kinetics. Systematically exploring how the relative rates of reaction and diffusion affect pattern formation -e.g. by varying the Damk\"ohler number— may help clarify the role of fluctuations in the emergence and stability of spatial structures.

\section*{Acknowledgments}
We acknowledge funding by the Basque Government through the BERC 2022-2025 program and by the Ministry of Science, Innovation and Universities: BCAM Severo Ochoa accreditation CEX2021-001142S/MICIN/AEI/10.13039/501100011033

\section*{Author Declarations}
The authors have no conflicts to disclose.

\section*{Author Contributions}
N.M. and M.E. designed the research project. M.E.F. performed the simulations and analysed the obtained data. The original draft of the manuscript was written by M.E.F and N.M. and all authors contributed to the final version of the manuscript. N.M. developed the numerical implementation. All the authors discussed and analyzed the results. All authors approved the final version of the manuscript.

\section*{Data Availability Statement}

The data that support the findings of this study are available within the article.


\appendix

\section{Analytical Solutions for Dirichlet and Neumann B.Cs}\label{app:Difftest1D}

For Dirichlet B.C given by 
\begin{equation}
C(x, t)=\frac{C_0 x}{L_x}+\sum_{n=1}^{\infty} \frac{2 C_0}{n \pi}(-1)^n \sin \left(\beta_n x\right) \exp \left(-D \beta_n^2 t\right),
\label{eq:dif1d_Dir_theo}
\end{equation}
where $\beta_n = n\pi/L_x$, $L_x = 20$, $C_0=1$ and $C(0, t) = 0$, $C(L_x , t) = 1$. For Neumann B.C, with an imposed flux $\Lambda = 0.01$ at $x = 0$ and a fixed value $C(L_x , t) = 0$ the analytical solution corresponds to 
\begin{align}
C(x, t) &= \Lambda(x - L_x) + \frac{4\Lambda}{L_x} \sum_{n=1,\text{odd}}^{\infty} \gamma_n^{-2} \sin^2 ({n\pi}/{4}) \nonumber \\
&\quad \times \cos(\gamma_n x) \exp\left(-D \gamma_n^2 t\right).
\label{eq:dif1d_Neum_theo}
\end{align}
where $\gamma_n= n\pi/2L_x$.

\section{Diffusional Tests Expanded}\label{app:Difftestsexpanded}
This section supports and extends the analysis of the interaction between deterministic and stochastic diffusion processes introduced in Section \ref{subsec:benchmarks-Dif}, focusing on their effects within a 2D square domain. The simulations were designed to isolate the contributions of thermal fluctuations and deterministic input diffusion across three distinct schemes.

\begin{figure}[!htbp]
\centering
\includegraphics[width=.5\textwidth]{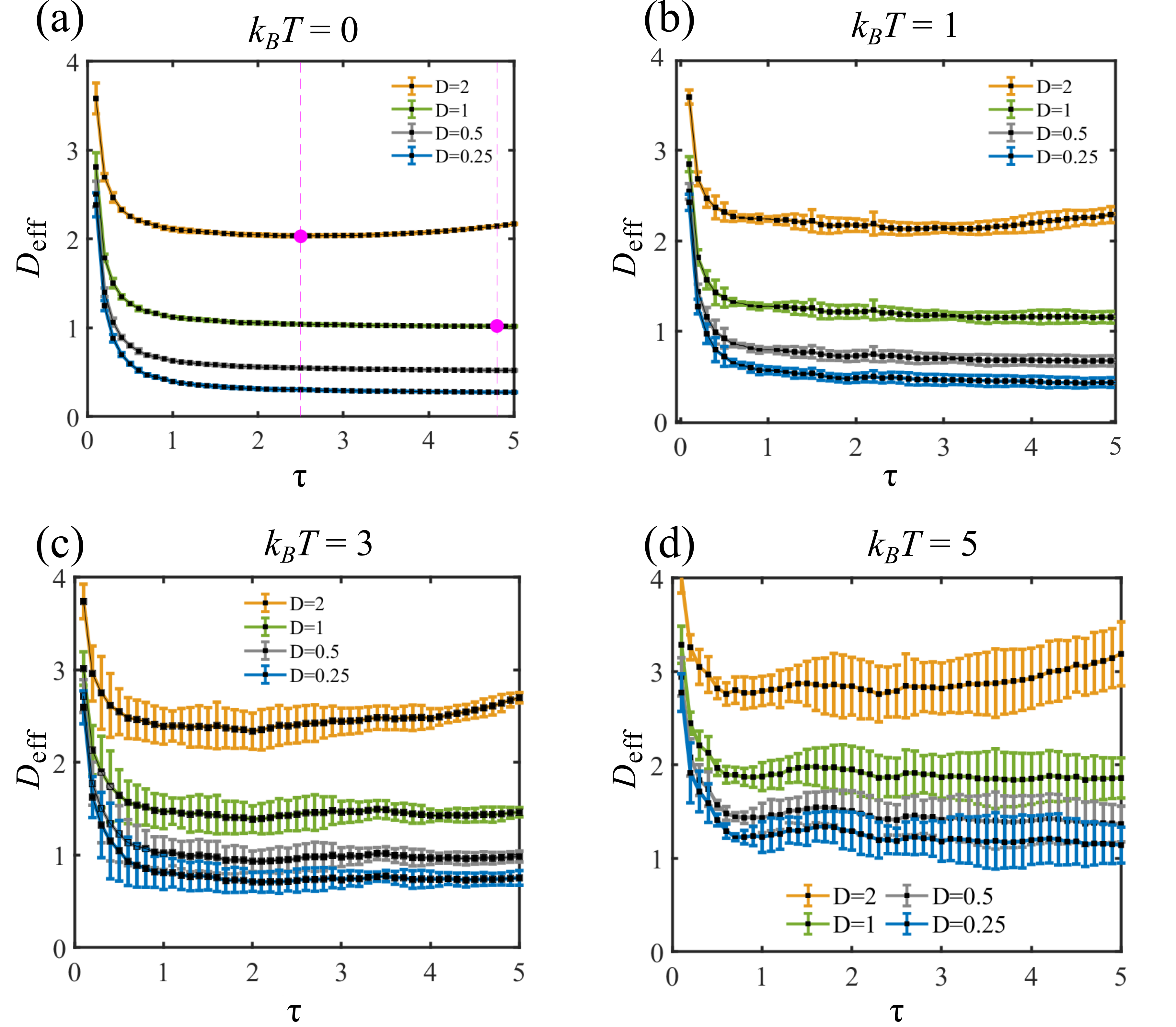}
\caption{Time evolution of measured diffusion coefficients in diffusional tests.
Subfigures (a)–(d) show $D_{\text{eff}}$ over time for $\kbt = 0,\ 1,\ 2,$ and $3$, respectively. In (a), magenta circles mark the times where $D_{\text{eff}}$ matches the input diffusion.}
\label{fig:RD_Deff_t_kbt}
\end{figure}

At $\kbt = 0$, Fig. \ref{fig:RD_Deff_t_kbt}a shows that particles with $D = 2$ reached the periodic boundaries within $t = 5\tau$, limiting the usable range for computing $D_{\text{eff}}^{\alpha}$. For $D = 2$ and $D = 1$, the measured values closely matched the input diffusion coefficients at $t = 2.47\tau$ and $t = 4.8\tau$, respectively, as indicated by the magenta circles. In contrast, for lower diffusivities ($D = 0.5$ and $D = 0.25$), the matching values were only reached after $t = 5\tau$, beyond the plotted range.

The time intervals used to compute $D_{\text{eff}}^{\alpha}$ for each $(D, \kbt)$ pair are listed in Table \ref{tab:RD_difftest_Temp_1_3_5}. They were chosen to capture the linear regime of concentration spread before boundary effects or nonlinearities appeared, ensuring consistent and reliable diffusivity estimates across all schemes.

\begin{figure*}[htbp]
\centering
\includegraphics[width=.97\textwidth]{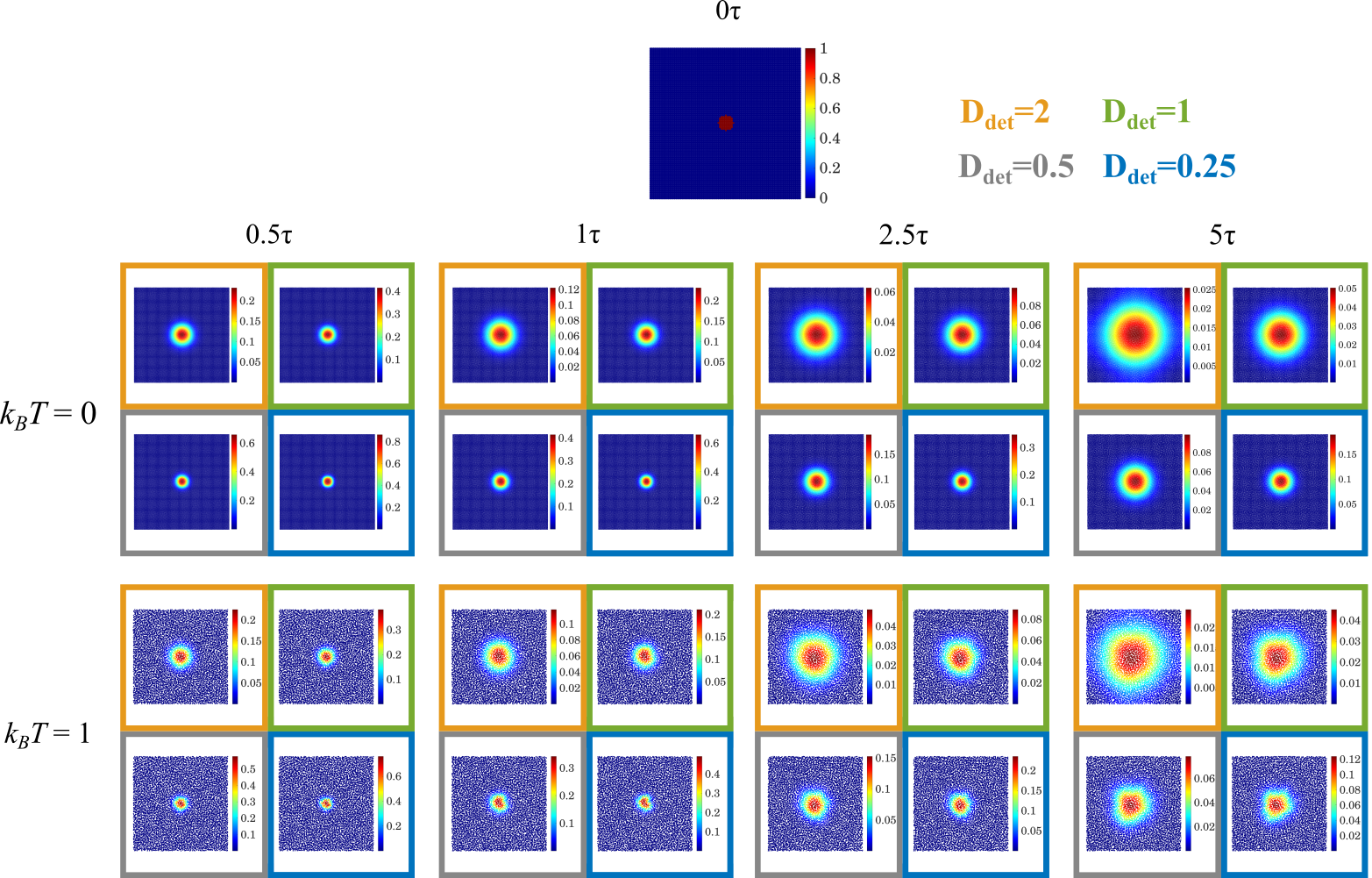}
\caption{Detailed scatterplots of species concentrations within the square domain, visualizing diffusion dynamics. The colored frames correspond to the diffusion input values, matching the color scheme used in Fig. \ref{fig:RD_Validation_Deff_3Sets} The colorbar highlights regions with low species concentrations, particularly evident for times $\tau>$1.}
\label{fig:RD_Diff_tests_scatter}
\end{figure*}

Fig. \ref{fig:RD_Diff_tests_scatter} offers a comparison of diffusion dynamics for \(\kbt = 0\) and \(\kbt = 1\), providing scatterplots of species concentrations within the domain. The colored frames corresponds to a distinct input diffusion value, using a color scheme consistent with Fig. \ref{fig:RD_Validation_Deff_3Sets}. These plots reveal how the stochastic component (evident at \(\kbt = 1\)) contributes to enhanced spreading compared to deterministic diffusion alone (\(\kbt = 0\)). The colorbar further highlights regions of low species concentration, which become particularly pronounced at longer times (\(\tau > 1\)). This visualization underscores the spatial implications of thermal effects, providing qualitative insight into the patterns observed in the quantitative analysis.  

Turning to the quantitative results, Table \ref{tab:RD_difftest_Temp_1_3_5} summarizes the effective diffusion coefficients (\(D_{\text{eff}}\)) and the associated estimation errors. Column 2 presents the deterministic diffusion estimates from Scheme 1, labeled as \(D_{\text{eff} | \kbt=0}\). Column 3 details the purely stochastic estimates from Scheme 2, denoted as \(D_{\text{eff} | \kbt=\{1, 3, 5\}, D_{\mathrm{det}}=0}\). Column 4 combines these effects, reporting \(D_{\text{eff} | \kbt > 0}\) from Scheme 3. Finally, column 5 captures the estimation errors between Schemes 1 and 2 when compared to Scheme 3. As the temperature (\(\kbt\)) increases, these errors grow, indicating the increasing difficulty of capturing \(D_{\text{eff}}\) with simplified assumptions.

\begin{table}[h!]
\caption{\label{tab:RD_difftest_Temp_1_3_5}%
Estimated effective diffusion coefficients $D_{\text{eff}}$ for different deterministic inputs $D_{\mathrm{det}}$ and thermal noise levels $\kbt=1,3,5$, with $\nu = 3.0$ fixed. Values for $D_{\text{eff}}$ are given with standard deviation and averaging interval in brackets. The additive approximation holds well ($<$2\%) only for $\mathrm{Sc} \gtrsim 0.24$ ($D_{\mathrm{det}} = 6.25$ and $12.5$) at $\kbt=0$.}
\begin{ruledtabular}
\scriptsize
\begin{tabular}{ccccc}
$D_{\mathrm{det}}$ & $D_{\text{eff}}|_{\kbt=0}$ & $D_{\text{eff}}|_{\kbt>0,\;D_{\mathrm{det}}=0}$ & $D_{\text{eff}}|_{\kbt>0}$ & Error [\%] \\
\hline
\multicolumn{5}{c}{$\kbt = 1$} \\
\hline
2     & 2.03 $\pm$ 0.009 &                            & 2.15 $\pm$ 0.074 [0.1–2 s] & 1.8 \\
1     & 1.02 $\pm$ 0.007 & \multirow{3}{*}{$0.16 \pm 0.105$ [0.1–10 s]} & 1.17 $\pm$ 0.065 [1–3 s]  & 1.0 \\
0.5   & 0.51 $\pm$ 0.005 &                                & 0.70 $\pm$ 0.058 [1–5 s]   & 2.9 \\
0.25  & 0.26 $\pm$ 0.005 &                                & 0.46 $\pm$ 0.051 [4–7 s]   & 7.5 \\
\hline
\multicolumn{5}{c}{$\kbt = 3$} \\
\hline
2     & 2.03 $\pm$ 0.009 &                            & 2.38 $\pm$ 0.123 [1–2 s]   & 6.7 \\
1     & 1.02 $\pm$ 0.007 & \multirow{3}{*}{$0.52 \pm 0.315$ [0.6–5 s]} & 1.44 $\pm$ 0.097 [1–3 s]  & 6.6 \\
0.5   & 0.51 $\pm$ 0.005 &                                & 0.97 $\pm$ 0.111 [1–5 s]   & 6.0 \\
0.25  & 0.26 $\pm$ 0.005 &                                & 0.74 $\pm$ 0.133 [1–5 s]   & 4.3 \\
\hline
\multicolumn{5}{c}{$\kbt = 5$} \\
\hline
2     & 2.03 $\pm$ 0.009 &                             & 2.78 $\pm$ 0.142 [0.5–1 s] & 9.1 \\
1     & 1.02 $\pm$ 0.007 & \multirow{3}{*}{$1.03 \pm 0.593$ [0.1–3 s]} & 1.91 $\pm$ 0.122 [0.5–2 s]& 6.9 \\
0.5   & 0.51 $\pm$ 0.005 &                                & 1.49 $\pm$ 0.149 [0.5–2 s] & 8.3 \\
0.25  & 0.26 $\pm$ 0.005 &                                & 1.26 $\pm$ 0.156 [0.5–3 s] & 3.4 \\
\end{tabular}
\end{ruledtabular}
\end{table}

\section{Kinetic Models Evaluated}\label{app:kinetics}

\subsubsection*{a. Simple Equilibrium Reaction}\label{subsubsec:S03}

The equilibrium reaction for the production of sulfur trioxide from sulfur dioxide and oxygen is described as \(2\mathrm{SO_2} + \mathrm{O_2} \rightleftharpoons 2\mathrm{SO_3}\). Measured  equilibrium concentrations\cite{atkins2010shriver} \([\mathrm{SO_2}] = 0.90~\text{M}\), \([\mathrm{O_2}] = 0.35~\text{M}\), and \([\mathrm{SO_3}] = 1.1~\text{M}\), lead to an equilibrium constant 
$K_{\mathrm{c}} = {{[\mathrm{SO_3}]^2}}/{{[\mathrm{SO_2}]^2 [\mathrm{O_2}]}} = 4.3\text{M}^{-1}$. Time is non-dimensionalized using the characteristic timescale of the dominant forward reaction: $\tau_{\text{r}} = 1 / (K_c [\mathrm{SO}_2]_0^2 [\mathrm{O}_2]_0)$, yielding $\tau_{\text{r}} \approx 0.82~\text{s}$. The source terms for each species \(\alpha\) are given by 
\begin{subequations}\label{eq:SO3_source-terms}
\begin{align}
    S^{\mathrm{SO}_2} &= -K_c \left[ 2 C^{\mathrm{SO}_2} C^{\mathrm{SO}_2} C^{\mathrm{O}_2} + 2 C^{\mathrm{SO}_3} C^{\mathrm{SO}_3} \right], \\
    S^{\mathrm{O}_2}  &= -K_c \left[ 2 C^{\mathrm{SO}_2} C^{\mathrm{SO}_2} C^{\mathrm{O}_2} + 2 C^{\mathrm{SO}_3} C^{\mathrm{SO}_3} \right], \\
    S^{\mathrm{SO}_3} &= -S^{\mathrm{SO}_2}.
\end{align}
\end{subequations}
Fig. \ref{fig:RD_Kinetictests_abc}a, shows that when the initial concentration of the species is at equilibrium, the system correctly preserves the steady condition. Whereas, when the initial concentration of $\mathrm{SO_3}$ is reduced, the system evolves toward a new equilibrium state that still satisfies $K_{\mathrm{c}} = 4.3$, consistent with theoretical predictions.\cite{stockmayer1944} 

\subsubsection*{b. Complex Reaction Kinetics}\label{subsubsec:PFR}  
For this benchmark, we consider the kinetics of four species --A, B, C, and D-- undergoing two elementary reactions: \(\mathrm{A} + 2\mathrm{B} \rightarrow \mathrm{C}\) and \(3\mathrm{C} + 2\mathrm{A} \rightarrow \mathrm{D}\), with the net source terms $S^{\alpha}$ for each species ${\alpha}$ given by \cite{UMichiganPFR} 
\begin{subequations}\label{eq:net_source_terms}
\begin{align}
S^{\mathrm{A}} &= -k_1 C^{\mathrm{A}} (C^{\mathrm{B}})^2 + \tfrac{2}{3} k_2 (C^{\mathrm{A}})^2 (C^{\mathrm{C}})^3, \\
S^{\mathrm{B}} &= -2k_1 C^{\mathrm{A}} (C^{\mathrm{B}})^2, \\
S^{\mathrm{C}} &= k_1 C^{\mathrm{A}} (C^{\mathrm{B}})^2 - k_2 (C^{\mathrm{A}})^2 (C^{\mathrm{C}})^3, \\
S^{\mathrm{D}} &= \tfrac{1}{3} k_2 (C^{\mathrm{A}})^2 (C^{\mathrm{C}})^3.
\end{align}
\end{subequations}
where the rate constants are \(k_1 = 10 \, \text{M}^{-2}/\mathrm{min}\) and \(k_2 = 15 \, \text{M}^{-4}/\mathrm{min}\). In Fig. \ref{fig:RD_Kinetictests_abc}b, we present the concentration evolution for a system with an initial equimolecular concentration, $F_{\mathrm{A_0}} = F_{\mathrm{B_0}} = 200 \, \mathrm{M}$. For this system, we use $\tau_{\text{r}} = 1 / (k_1 [\mathrm{A}_0] [\mathrm{B}_0]^2)$. Overall, from Fig. \ref{fig:RD_Kinetictests_abc}b, we can observe an excellent agreement between \textcolor{black}{SDPD} simulations and the corresponding ODE evolution for all the species concentrations.

\subsubsection*{c. Michaelis-Menten Reaction Model}\label{subsubsec:MichMenten}

We now consider the enzymatic conversion of a substrate ($\mathrm{S}$) into a product ($\mathrm{P}$) via the classical Michaelis–Menten mechanism. The process involves two reversible and irreversible steps: the reversible formation of an enzyme–substrate complex ($\mathrm{E} + \mathrm{S} \rightleftharpoons \mathrm{ES}$) followed by the irreversible production of product ($\mathrm{ES} \rightarrow \mathrm{E} + \mathrm{P}$). The enzyme is present in two forms: the free enzyme ($\mathrm{E_0}$) and the bound complex ($\mathrm{E_1} \equiv \mathrm{ES}$). The corresponding source terms $S^\alpha$ for each species are defined by the following set of ordinary differential equations (ODEs):
\begin{subequations}\label{eq:MichMentenODEs}
\begin{align}
S^{\mathrm{S}} &= -k_1 C^{\mathrm{S}} C^{\mathrm{E}_0} + k_{\tiny -1} C^{\mathrm{E}_1}, \\
S^{\mathrm{E}_0} &= -k_1 C^{\mathrm{S}} C^{\mathrm{E}_0} + k_{\tiny -1} C^{\mathrm{E}_1} + k_2 C^{\mathrm{E}_1}, \\
S^{\mathrm{E}_1} &= k_1 C^{\mathrm{S}} C^{\mathrm{E}_0} - k_{\tiny -1} C^{\mathrm{E}_1} - k_2 C^{\mathrm{E}_1}, \\
S^{\mathrm{P}} &= k_2 C^{\mathrm{E}_1}.
\end{align}
\end{subequations}
Equations \eqref{eq:MichMentenODEs} define the source terms \(S^{\alpha}\) for each concentration field in the Michaelis-Menten system. For the current study, we used the following values: \(k_1 = 10 \text{M}^{-1} \text{s}^{-1}\), \(k_{\tiny -1} = 1 \text{M}^{-1} \text{s}^{-1}\), and \(k_2 = 5 \text{M}^{-1} \text{s}^{-1}\), with initial concentrations of \(\mathrm{S} = 1\) M and \(\mathrm{E_0} = 0.08\) M. The characteristic time is defined as $\tau_{\text{r}} = 1 / (k_{\mathrm{ref}} [\mathrm{S}_0])$, with $k_{\mathrm{ref}} = 10\,\mathrm{M}^{-1}\mathrm{s}^{-1}$ and $[\mathrm{S}_0] = 1\,\mathrm{M}$, yielding $\tau_{\text{r}} = 0.1\,\mathrm{s}$.  The concentration profiles in Figure \ref{fig:RD_Kinetictests_abc}c align with theoretical predictions,\cite{Qian2021} validating the \textcolor{black}{SDPD} method's accuracy in capturing Michaelis-Menten dynamics. 

\section{Analytical Solution Advection-Diffusion flow in a Channel}\label{app:advdiff}

Symmetry about \(y = 0\) (Fig.~\ref{fig:RD_2d_Poiseuille}a) allows the analysis to focus on the half-channel region \((0 \leq y \leq H)\), applying the symmetry boundary condition \(\der C / \der y \big|_{y=0} = 0\) to simplify the domain. The analytical solution for this configuration, derived by Kuzmin et al.,\cite{kuzmin2013lattice} is expressed as
\begin{equation}
\footnotesize{
    C(x, y) = C_{\mathrm{w}} \left[1 - \sum_m C_m \mathrm{e}^{-\frac{m^4 x}{H \mathrm{Pe}} - \frac{m^2 y^2}{2 H^2}} \, {}_1 F_1 \left(-\frac{m^2}{4} + \frac{1}{4}, \frac{1}{2}, \frac{m^2 y^2}{H^2}\right)\right],
    }
    \label{eq_RD_ADE_analyticsol}
\end{equation}
where \({}_1 F_1\) is the hypergeometric function. The parameter \(m\) represents the roots satisfying \(\small{{}_1 F_1 \left({1}/{4} - {m^2}/{4}, {1}/{2}, m^2 \right)=0}\). The coefficients \(C_m\) are obtained through integrals of the hypergeometric function, yielding the full solution formulation:

\begin{equation}
\small{
C_m = -C_{\mathrm{w}} \frac{\int_0^1 \left(1 - \xi^2\right) \mathrm{e}^{-{m^2 \xi^2}/{2}} \, {}_1 F_1\left(-\frac{m^2}{4} + \frac{1}{4}, \frac{1}{2}, m^2 \xi^2\right) \mathrm{d} \xi}{\int_0^1 \left(1 - \xi^2\right) \mathrm{e}^{-m^2 \xi^2} {}_1 F_1\left(-\frac{m^2}{4} + \frac{1}{4}, \frac{1}{2}, m^2 \xi^2\right)^2 \mathrm{d} \xi}.
}
\label{eq_RD_ADE_1F1_hypergeom-integrals}
\end{equation} 

\section{Reaction–Diffusion Model Equations}
\label{app:ReacDiffEquations}

This appendix provides the full formulation of the reaction–diffusion (RD) system used in the Turing-pattern formation benchmark (Section \ref{subsec:TuringPattern}). The model involves two interacting chemical species: an activator $C^{Ac}(t,x,y)$ and an inhibitor $C^{In}(t,x,y)$. Their evolution in space and time is governed by nonlinear reaction kinetics and diffusion, following the system originally proposed by Cooper and co-authors.\cite{cooper2018ancient} The governing equations read:

\begin{align}
\frac{\der C^{Ac}}{\der t} &= S^{Ac}(C^{Ac}, C^{In}) - k_{\mathrm{deg}}^{Ac}\,C^{Ac} + D^{Ac}\,\Delta C^{Ac}, \\[1em]
\frac{\der C^{In}}{\der t} &= S^{In}(C^{Ac}, C^{In}) - k_{\mathrm{deg}}^{In}\,C^{In} + D^{In}\,\Delta C^{In}.
\end{align}

Here, $D^{Ac}$ and $D^{In}$ are the diffusion coefficients of the activator and inhibitor, while $k_{\mathrm{deg}}^{Ac}$ and $k_{\mathrm{deg}}^{In}$ are their respective degradation rates. The source terms $S^{Ac}(C^{Ac}, C^{In})$ and $S^{In}(C^{Ac}, C^{In})$ encode the nonlinear kinetics and are defined piecewise as:

{\footnotesize
\begin{subequations} \label{eq:Turing-Suv}
\begin{align}
S^{Ac}(C^{Ac}, C^{In}) &= \begin{cases}
0, & \text{if } k_1^{Ac} C^{Ac} + k_2^{Ac} C^{In} + k_3^{Ac} < 0, \\
S_{\max}^{Ac}, & \text{if } k_1^{Ac} C^{Ac} + k_2^{Ac} C^{In} + k_3^{Ac} > S_{\max}^{Ac}, \\
k_1^{Ac} C^{Ac} + k_2^{Ac} C^{In} + k_3^{Ac}, & \text{otherwise},
\end{cases} \label{eq:Turing-Su} \\[1em]
S^{In}(C^{Ac}, C^{In}) &= \begin{cases}
0, & \text{if } k_1^{In} C^{Ac} + k_2^{In} C^{In} + k_3^{In} < 0, \\
S_{\max}^{In}, & \text{if } k_1^{In} C^{Ac} + k_2^{In} C^{In} + k_3^{In} > S_{\max}^{In}, \\
k_1^{In} C^{Ac} + k_2^{In} C^{In} + k_3^{In}, & \text{otherwise}.
\end{cases} \label{eq:Turing-Sv}
\end{align}
\end{subequations}
}

The kinetic parameters are non-dimensionalized consistently with the chosen scalings: time is scaled by the inhibitor diffusion timescale, $t^* = l^2 / D^{In}$, and concentrations are implicitly scaled by a reference activator concentration $C^* = 1\,\text{nM}$. As a result, all rate constants $k_i^\alpha$ and degradation rates $k_{\mathrm{deg}}^\alpha$ are expressed in units of inverse time and are dimensionless within this framework.

The initial conditions are defined as:

\begin{subequations} \label{eq:Turing-IC}
\begin{align}
C^{Ac}(0, x, y) &=
\begin{cases}
C_0^{Ac}, & \text{if } \left(x - x_i\right)^2 + \left(y - y_i\right)^2 < R_{\mathrm{spot}}^2, \\
& \quad \text{for } i \in \{0, \dots, n_{\mathrm{spot}} - 1\}, \\
0, & \text{otherwise},
\end{cases} \\[0.5em]
C^{In}(0, x, y) &= 0.
\end{align}
\end{subequations}

The model parameters are chosen to favor pattern formation, following the conditions reported by Miura et al.\cite{Miura2004} The values used in this study are:
$k_1^{Ac} = 0.08$, $k_2^{Ac} = -0.08$, $k_3^{Ac} = 0.04$, $k^{Ac}_{\text{deg}} = 0.03$, $S_{\max}^{Ac} = 0.2$, $D^{Ac} = 0.02$, $k_1^{In} = 0.16$, $k_2^{In} = 0$, $k_3^{In} = -0.05$, $k^{In}_{\text{deg}} = 0.08$, $S_{\max}^{In} = 0.5$, and $D^{In} = 0.6$.

The initial activator concentration ($C_0^{Ac} = 5$) is applied within $n_{\mathrm{spot}} = 6$ circular spots of radius $R_{\mathrm{spot}} = 4.5$, aligned along the line $y = L_s/2$ and centered at $x_i \in {0, L_s/5, 2L_s/5, 3L_s/5, 4L_s/5, L_s}$, with $L_s=75\ulen$.

\section{Estimation of $D_{\text{eff}}$ via MSD Fitting}
\label{appB:MSD}

To evaluate whether the additive relation $D_{\text{eff}}^{\alpha} \approx D_{\text{det}}^{\alpha} + D_{\text{stoch}}^{\alpha}$ holds in the context of Turing pattern formation, we compared two simulation setups of the activator–inhibitor system.

In the first setup, a hybrid stochastic–deterministic configuration was used, with both thermal fluctuations ($\kbt = 0.025$) and nonzero deterministic diffusivity ($D^\alpha_{\text{det}} > 0$). In the second, an athermal configuration ($\kbt = 0$) was used, where the effective diffusion coefficient was constructed as $D^\alpha_{\text{eff}} = D^\alpha_{\text{det}} + \tilde{D}_{\text{det}}$. The correction $\tilde{D}_{\text{det}}$ was chosen to match the contribution of thermal fluctuations to diffusion, and was estimated via a reference simulation with $D^\alpha_{\text{det}} = 0$ and $\kbt = 0.025$.

\begin{figure}[!h]
\centering
\includegraphics[width=.49\textwidth]{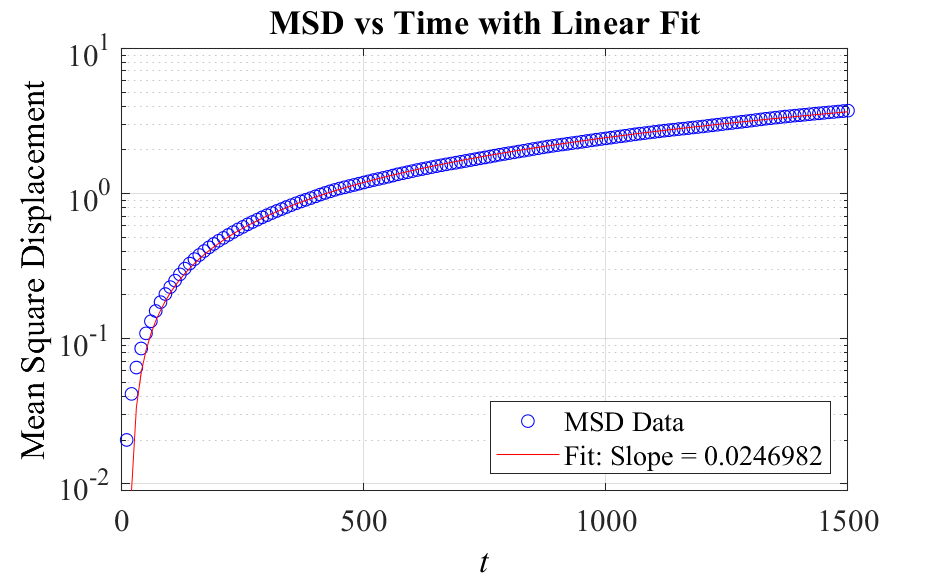}
\caption{Mean Square Displacement (MSD) as a function of timestep for particles with zero input diffusion at \(\kbt = 0.025\). A linear fit to the data provides the diffusion coefficient \(D\), derived from the slope of the MSD curve.}
\label{fig:RD_MSD}
\end{figure}

In this reference \textcolor{black}{SDPD} simulation, we computed the mean squared displacement (MSD) from particle trajectories. As shown in Fig. \ref{fig:RD_MSD}, the MSD increases linearly with time, consistent with diffusive motion driven by thermal fluctuations. A linear fit yields $\text{MSD}(t) \approx 0.0247\, t$. Using the 2D Einstein relation $\text{MSD}(t) = 4D\, t$, we obtain $\tilde{D}_{\text{det}} \approx 0.00625\, \ulen^2/\tau$.

This value, $\tilde{D} = \tilde{D}_{\text{det}}$ is used in the main text to construct purely deterministic simulations that match the effective diffusivity (${D}_{\text{eff}}$) of their stochastic counterparts.

\section{Supplementary Tables}\label{app:SItables}
\begin{table}[htbp]
\caption{\label{tab:sdpd-params}Units and simulation parameters used in the SDPD models.}
\label{tab:units_params}
\begin{ruledtabular}
\begin{tabular}{ll}
\multicolumn{2}{c}{\textbf{Basic Units}} \\
\hline
Length & $\ulen$ \\
Energy & $k_B T$ \\
Mass & $m_f$ \\
\hline
\multicolumn{2}{c}{\textbf{Derived Units}} \\
\hline
Time & $\tau = \ulen \sqrt{\frac{m_f}{k_B T}}$ \\
Diffusion constant & $D = \frac{ \ulen^2}{t_0} =  \ulen \sqrt{\frac{k_B T}{m_f}}$ \\
Kinematic viscosity & $\nu = \frac{ \ulen^2}{t_0} =  \ulen \sqrt{\frac{k_B T}{m_f}}$ \\
Viscosity & $\eta = \frac{m_f}{t_0  \ulen} = \frac{\sqrt{m_f k_B T}}{ \ulen^2}$ \\
\end{tabular}
\end{ruledtabular}

\vspace{1em}

\begin{ruledtabular}
\begin{tabular}{lc}
\textrm{Parameter} & \textrm{Value (non-dimensional)} \\
\colrule
Mass density, $\rho$ & 1.0\\
Dynamic viscosity, $\eta$ & 3.0 \\
Thermal energy, $k_B T$ &  0--5\\
Speed of sound, $c_0$ & 50 \\
Time step, $\Delta \tau$ & $10^{-3}\,\tau$ \\
Reference length, $l$ & 1 (unit length) \\
Reference time, $\tau$ & 1 (unit time) \\
Simulation box length & $l_{\text{box}}, L_s$ (varied)\\
Initial lattice spacing, $dx$ & 0.2 \\
Cutoff radius, $h$ & 4\,$dx$ \\
\end{tabular}
\end{ruledtabular}
\end{table}

%
    
\end{document}